\documentclass{PoS}

\title{Lattice Landau Gauge and Algebraic Geometry}

\ShortTitle{Lattice Landau Gauge and Algebraic Geometry}

\author{\speaker{Dhagash Mehta}$^{1,2,3}$\\
E-mail: \email{dhagash.mehta@nuim.ie}}
\author{Andre Sternbeck$^1$\thanks{New address since October 2009:
Institut f\"ur Theoretische Physik, Universit\"at Regensburg, D-93040
Regensburg, Germany}, Lorenz von Smekal$^{1,4}$, Anthony G Williams$^1$ \\
        \llap{$^1$} Centre for the Subatomic Structure of Matter, School of Chemistry and Physics, University of Adelaide, SA 5005, Australia.\\
        \llap{$^2$} Department of Mathematical Physics, National University of Ireland Maynooth, Maynooth, Co. Kildare, Ireland.\\
        \llap{$^3$} Department of Physics, Imperial College London, Prince Consort Road, London SW7 2AZ, UK.\\
        \llap{$^4$} Institut für Kernphysik, Technische Universität Darmstadt, Schlossgartenstr. 9, 64289 Darmstadt, Germany.}

\abstract{Finding the global minimum of a multivariate function
  efficiently is a fundamental yet difficult problem in many branches of theoretical
  physics and chemistry. However, we observe that there are many physical systems for
  which the extremizing equations have polynomial-like non-linearity. This
  allows the use of Algebraic Geometry techniques to solve these
  equations completely. The global minimum can then
  straightforwardly be found by the second derivative test.
  As a warm-up example, here we study lattice Landau gauge for compact
  U(1) and propose two methods to solve the corresponding gauge-fixing
  equations. In a first step, we obtain  \emph{all} Gribov copies on one
  and two dimensional lattices. For simple $3 \times 3$ systems their
  number can already be of the order of thousands. We anticipate that the
  computational and numerical algebraic geometry methods employed
  have far-reaching implications beyond the simple but illustrating
  examples discussed here.}

\FullConference{International Workshop on QCD Green's Functions,\\
Confinement, and Phenomenology - QCD-TNT09\\
		 September 07 - 11 2009\\
		 ECT Trento, Italy}

\begin{document}

%====================================================================================
%====================================================================================

\section{Introduction}
Finding the global minimum of a multivariate function is one of the
most important tasks in statistical mechanics, condensed matter
theory, lattice gauge theories and theoretical chemistry. Most of
the usual methods to minimize a multivariate function are based on the
Newton-Raphson method, where a start solution is guessed and then is
refined by successive iterations. There are several efficient
refinements of this minimization procedure, e.g., simulated
annealing. However, they are often not successful as they can
get trapped in a local minimum instead of the global one. This
problem introduces an error of an \textit{unknown} order.

A typical example is the Landau gauge on the lattice:  in the continuum the corresponding Landau gauge fixing condition $\partial_{\mu} A_{\mu} = 0$ is solved to fix the gauge, whereas on the lattice, usually the so-called standard lattice Landau gauge-fixing (LLG) functional (whose first derivatives with respect to the gauge parameters are the corresponding gauge-fixing conditions on the lattice) is numerically minimized~\cite{Iritani:2009mp,Bogolubsky:2009dc,Cucchieri:2007md,Leinweber:1998uu}. We take the example of the LLG functional for compact U($1$), which is given by
\begin{equation}
 F = \sum_{i,\mu}(1-\cos (\phi_{i,\mu} + \theta_{i+\hat{\mu}} - \theta_{i})),\label{eq_hamiltonian_llg_compact_u_1}
\end{equation}
where the $\theta_{i}\in (-\pi,\pi]$ variables sit on the $i$th
lattice site of a $d$-dimensional lattice grid, $\hat{\mu}$ is the
unit vector in $\mu$ direction, and the $\phi_{i,\mu}\in (-\pi,\pi]$
variables sit on the links between the $i$th and $(i+\hat{\mu})$th
lattice sites. Here the $\phi$-variables are related to the gauge
potential in the corresponding continuum gauge theory and the
$\theta$-variables are the gauge transformations with respect to which
$F$ is minimized. The special case when all $\phi_{i,\mu}$ are zero is
called the trivial orbit. We will also frequently use random choices
for the link angles $\phi_{i,\mu}$ which we then refer to as
random orbits. The Hessian matrix of this functional is the
Faddeev-Popov operator and its determinant is known as the
Faddeev-Popov determinant~\cite{Faddeev:1967fc} on the lattice.

In order to compare the results from lattice Landau gauge to those
from functional methods in the continuum such as studies of
Dyson-Schwinger equations~\cite{Alkofer:2000wg}, one should in
principle find all stationary points of $F$ under variation with
respect to the gauge variables $\theta_{i}$
\cite{vonSmekal:2007ns,vonSmekal:2008es,vonSmekal:2008ws,Mehta:2009},
that is {\em all} solutions to the corresponding LLG equations on the
lattice, %which are the first derivatives of $F$ equated to zero,
\begin{equation}
 \frac{\partial F}{\partial \theta_i} = 0 \; .\label{eq:dF_by_dtheta}
\end{equation}
The solutions to these equations are the lattice analogues of Gribov
copies~\cite{Gribov:1977wm}.\footnote{For recent progress in the
continuum, see also ~\cite{Ilderton:2007dh, Ilderton:2007qy}.} The
sum over all those Gribov copies weighted by the sign of the
Faddeev-Popov determinant evaluated at each copy would then be
independent of the gauge orbit by the Poincar\'e-Hopf theorem and this
could thus be used to define the measure of Landau gauge on the
lattice analogous to standard Faddeev-Popov theory with
Becchi-Rouet-Stora-Tyutin (BRST) symmetry.

The reason why this procedure fails is that the signs of all Gribov
copies exactly cancel one another, always yielding an exact zero for
their sum which computes the vanishing Euler characteristic of the
lattice gauge group manifold in the standard implementation of the
Landau gauge. This zero sum is the origin for the famous Neuberger
$0/0$ problem of standard BRST formulations on the lattice
\cite{Neuberger:1986vv,Neuberger:1986xz}, which is the statement that
the expectation value of any gauge-invariant observable is of
indefinite form $0/0$ in such a formulation.

The $0/0$ problem can be overcome by stereographically projecting the
lattice gauge group onto a manifold whose Euler characteristic is
non-zero \cite{vonSmekal:2008ws}. This provides a well-defined lattice BRST
formulation on the projected manifold and it can serve as a
non-perturbative definition of BRST symmetry in the continuum limit.

The Gribov copies of compact U($1$) are of course lattice
artifacts. However, the gauge group is a direct product of
odd-dimensional and compact manifolds (circles in this case), as it is for
every SU($N$) gauge theory on the lattice also. This is sufficient to
conclude that it also shares the same $0/0$ problem with SU($N$)
theories.  Moreover, it turns out that the
Neuberger $0/0$ problem in SU($N$) is avoided when that of its maximal
Abelian subgroup $U(1)^{N-1}$ is \cite{Schaden:1998hz}, because the
coset manifold consist of even dimensional spheres whose Euler
characteristic is 2. It is therefore obviously important to understand
the Gribov copy and Neuberger problems in compact U($1$) gauge models.

At the same time, for the trivial orbit the functional $F$ in
Eq.~(\ref{eq_hamiltonian_llg_compact_u_1}) also represents the
Hamiltonian of an XY model which stands alongside the Ising and
Heisenberg models as one of the most intensely studied systems in
statistical mechanics and condensed matter theory.
For a random orbit, it is that of a random phase XY model
(RPXYM). These have been widely used as simple models of classical
superconductivity, to study aspects of high-$T_{c}$ superconductors,
to describe the XY magnet with random Dzyaloshinski-Moriya
interactions or a positional disordered Josephson junction array, to
name a few.  In either case, the calculations (related to Domain Wall
Renormalization Group studies) boil down to obtaining the global
minimum of the Hamiltonian with respect to the $\theta$-variables for
a given set of $\phi_{i,\mu}$'s (see, e.g., \cite{AK:02}).
In Ref.~\cite{NH:95}, it was shown in lower dimensional models that
one can first perform a duality transformation then
followed by a numerical minimization so that a minimization algorithm has to
search only within the space of local minima. Recently, Akino and
Kosterlitz~\cite{AK:02} have used this method to obtain the global
minimum of a RPXYM Hamiltonian. This method, while giving more
confidence in the final results with less numerical effort, still
relies on numerical algorithms which are known to fail for larger
systems.

The knowledge of all stationary points is important in recent
studies of the potential energy landscape of the classical statistical
mechanical systems such as the RPXYM and its relation to phase
transitions~\cite{pettini:04, Kastner:2008zz, Wales:04}.
An effort to numerically classify all Gribov copies of compact U(1) in
2 dimensions was done in~\cite{deForcrand:1994mz}.

Here, we take up the general problem of finding all solutions to
the standard LLG equations for compact U($1$) or the stationary points
of the corresponding XY model. The corresponding equations are
non-linear which makes them difficult to deal with analytically or
numerically. We observe, however, that this non-linearity is
\textit{polynomial-like} and the equations can be transformed into
systems of polynomial equations. Then we can use Algebraic Geometry
techniques to solve these systems and obtain all stationary points of
the corresponding function and in particular the global minimum
accurately. This procedure is applicable to a wide range of
interesting physical problems such as obtaining all stationary points
of the classical Hamiltonians of Ising and Heisenberg models; solving
classical field equations for pure lattice gauge field theories,
Abelian Higgs model, etc.; finding a generic form of SU($N$) matrices;
obtaining the vacua of Supersymmetric potentials; and
many more (see Ref.~\cite{Mehta:2009}). The corresponding gauge-fixing equations
for the modified LLG, proposed in Refs.~\cite{vonSmekal:2007ns,vonSmekal:2008es}, for the compact U($1$) case can also be shown to have polynomial-like non-linearity but their structure is more complicated~\cite{Mehta:2009}. Here, we restrict the discussion to the standard LLG.

In the one-dimensional lattice case, the LLG equations have been
exactly solved for both periodic and anti-periodic boundary conditions
elsewhere~\cite{Mehta:2009,vonSmekal:2007ns,vonSmekal:2008es} where
the corresponding equations are \textit{treated} as systems of linear
equations. However, for higher dimensional lattices the corresponding
equations are \textit{genuinely} non-linear and the same method can
not be used. We will show that the systems of these non-linear
equations can be viewed as systems of multivariate polynomial
equations and therefore can be described by the language of Algebraic
Geometry. We will then discuss two promising approaches to solve the
polynomial systems: the numerical polynomial homotopy continuation
(NPHC) method and the computational Algebraic Geometry approach (in
the Appendix) and present our results. Though our goal is to solve the
gauge fixing conditions for higher dimensional lattices, we will
explain the methods using the one-dimensional case because: (1) we
already know the exact solutions for this case and so the results from
the new approaches will have a precise comparison, (2) the
one-dimensional case, interpreted as an Algebraic Geometry problem,
provides all the essence of the higher dimensional generalizations.

\section{The Extremizing Equations as Polynomial Equations}
Here, we show that the problem of solving the extremizing equations in terms of the $\theta$-variables can be transformed into that of solving a system of multivariate polynomial equations. We first note that, for a one-dimensional lattice having $n$ lattice sites, the directional index $\mu$ is irrelevant and $\hat{\mu}$ is just $1$. Thus, the corresponding extremizing equations are
\begin{equation}
f_{i}(\phi^{\theta}) = \sin(\phi_{i}^{\theta}) - \sin(\phi_{i-1}^{\theta}) = 0,
\end{equation}
where $\phi^{\theta}_{i} = (\phi_{i} + \theta_{i+1} - \theta_{i}) \in (-\pi,\pi]$, for all $i=1,\dots ,n$. These are the gauge-fixing equations (\ref{eq:dF_by_dtheta}) on the lattice and reproduce the continuum Landau gauge ($\partial_{\mu} A_{\mu} = 0$) in the naive continuum limit. These equations are also the steady state equations of a variant of the famous Kuramoto model to study synchronization in mathematical biology~\cite{Acebron:2005zz}. It can be shown that by using anti-periodic boundary conditions on both $\phi$- and $\theta$-variables the global gauge freedom can be completely fixed~\cite{vonSmekal:2007ns,vonSmekal:2008es,Mehta:2009}. While dealing with periodic boundary conditions one can get rid of the residual gauge freedom by taking, say, $\theta_{n}$ to be zero and removing the equation $f_{n}=0$ from the system~\cite{Mehta:2009}. For simplicity, we take the trivial orbit case (i.e., all $\phi_{i}=0$). Using the trigonometric identity, $\sin (x + y) = \sin x \cos y + \sin y \cos x$ and writing $\cos \theta_{i} \equiv c_{i}$ and $\sin \theta_{i} \equiv s_{i}$, we get
\begin{equation}
f_{i}(c,s) = c_{i}(s_{i+1} + s_{i-1}) - s_{i}(c_{i+1} + c_{i-1}) = 0\; , \label{eq:one_dim_standard_trig_to_poly}
\end{equation}
for all $i=1,\dots ,n$. This is merely a change of notation. However, we add additional equations in the system for each site $i$, namely,
\begin{equation}
g_{i}(c,s) = s_{i}^2 + c_{i}^2 -1 = 0,\label{eq:one_dim_standard_general_poly}
\end{equation}
for all $i=1,\dots ,n$. Now, the combined system of all $f_{i}(c,s)$ and $g_{i}(c,s)$ is not just a change of notation but all the $c_{i}$ and $s_{i}$ are algebraic variables and the equations are multivariate polynomial equations, i.e., the fact that $c_{i}$ and $s_{i}$ are originally $\sin \theta_{i}$ and $\cos \theta_{i}$ is taken care of by the constraint equations (\ref{eq:one_dim_standard_general_poly}). In general, for the one-dimensional lattice with $n$ lattice sites, we have in total $2 n$ polynomial equations and $2 n$ variables.

\section{Numerical Polynomial Homotopy Continuation Method}
In the Appendix, a method to exactly solve systems of multivariate equations, called the Groebner basis technique, and its current status are briefly explained. Here we discuss a numerical method to solve a system of multivariate equations, called the numerical polynomial homotopy continuation (NPHC) method. To explain the NPHC method let us consider a system of multivariate polynomial equations, say $P(x)=0$, where $P(x) = (p_{1}(x),\dots ,p_{m}(x))^{T}$ and $x = (x_{1},\dots ,x_{m})^{T}$, that is \textit{known to have isolated solutions}, e.g., the above mentioned LLG equations after eliminating the global gauge freedom. Now, there is a classical result, called the \textit{Classical Bezout Theorem}, that asserts that for a system of $m$ polynomial equations in $m$ variables, for generic values of coefficients, the maximum number of solutions in $\mathbb{C}^{m}$ is $\prod_{i=1}^{m}d_{i}$, where $d_{i}$ is the degree of the $i$th polynomial. This bound, called the classical Bezout bound (CBB), is exact for generic values (i.e., roughly speaking, non-zero random values) of coefficients, e.g., for the one-dimensional LLG equations with $n$ number of lattice sites and with periodic boundary conditions, this number is $2^{2 n}$ (because there are $2n$ polynomials each of which is a degree $2$ polynomial). The \textit{genericity} is well-defined and the interested reader is referred to Ref.~\cite{SW:95,Li:2003} for details.

Using the so-called Bernstein-Khovanskii-Kushnirenko theorem, a tighter bound, which takes the sparsity of the system into account, can be computed. This bound is called the BKK root count or the mixed volume (or stable mixed volume (SMV) \cite{Li:2003,Li96thebkk,HB:97} in general). The reader is referred to the above references to get a precise definition of the mixed volume and its computation.

Based on either of these bounds on the number of complex solutions, a \textit{homotopy} can be constructed as
\begin{equation}
 H(x, t) = \gamma (1 - t) Q(x) + t\; P(x),
\end{equation}
where $\gamma$ is a random complex number. $Q(x) = (q_{1}(x),\dots ,q_{m}(x))^{T}$ is a system of polynomial equations with the following properties: (1) the solutions of $Q(x) = H(x,0)=0$ are known or can be easily obtained. $Q(x)$ is called the \textit{start system} and the solutions are called the \textit{start solutions},
(2) the number of solutions of $Q(x) = H(x,0) = 0$ is equal to the CBB or SMV for $P(x)=0$, (3) the solution set of $H(x,t)=0$ for $0\le t \le 1$ consists of a finite number of smooth paths, called homotopy paths, each parameterized by $t\in [0,1)$, and (4) every isolated solution of $H(x,1)=P(x)=0$ can be reached by some path originating at a solution of $H(x,0)=Q(x)=0$. One can then track all the paths corresponding to each solution of $Q(x)=0$ from $t=0$ to $t=1$ and reach $P(x)=0=H(x,1)$. Here, a randomly chosen complex $\gamma$ ensures the paths are well-behaved. By implementing an efficient path tracker algorithm, all isolated solutions of a system of multivariate polynomials system can be obtained.

To illustrate how the method works, we take a univariate polynomial from Ref.~\cite{SW:95} to make this discussion self-consistent, say $z^{2}-5 = 0$, pretending that we do not know its solutions, i.e., $z=\pm \sqrt{5}$. We first define a family of problems as
\begin{equation}
 H(z,t) = (1-t)(z^{2}-1) + t(z^{2}-5) = z^{2} - (1 + 4 t) = 0, \label{eq:one_var_homotopy}
\end{equation}
where $t\in [0,1]$ is a parameter. For $t=0$, we have $z^{2}-1 = 0$ and at $t=1$ we recover our original problem. The problem of getting all solutions of the original problem now reduces to tracking solutions of $H(z,t)=0$ from $t=0$ where we know the solutions, i.e., $z=\pm 1$, to $t=1$. The choice of the piece $z^{2}-1$ in Eq.~(\ref{eq:one_var_homotopy}), called the \textit{start system}, should be clear now: this system has the same number of solutions as the CBB of the original problem and is easy to solve. Now, one of the ways to track the paths is to solve the differential equation that is satisfied along all solution paths, say $z_{i}^{*}(t)$ for the $i$th solution path:
\begin{equation}
 \frac{dH(z_{i}^{*}(t),t)}{dt} = \frac{\partial H(z_{i}^{*}(t),t)}{\partial z} \frac{d z_{i}^{*}(t)}{dt} + \frac{\partial H(z_{i}^{*}(t),t)}{\partial t} = 0.
\end{equation}
This equation is called the Davidenko differential equation. Inserting (\ref{eq:one_var_homotopy}) into this equation, we have
\begin{equation}
 \frac{d z_{i}^{*}(t)}{dt} = -\frac{2}{z_{i}^{*}(t)}.
\end{equation}
This initial value problem can be solved numerically (again, pretending that an exact solution is not known) with the initial conditions as $z_{1}^{*}(0) = 1$ and $z_{2}^{*}(0) = -1$. The other approach is to use Euler's predictor and Newton's corrector methods. We do not intend to discuss the actual path tracker algorithm used in practice, but it is important to mention that in the path tracker algorithms used in practice, almost all apparent difficulties have been resolved, such as tracking singular solutions, multiple roots, solutions at infinity, etc. For the sake of completeness, we should also mention here that in the actual path tracker algorithms the homotopy is randomly complexified to avoid singularities by $\gamma$, called the \textit{gamma trick}, i.e., taking
\begin{equation}
 H(z,t) = \gamma (1-t)(z^{2}-1) + t(z^{2}-5) = 0.
\end{equation}
There are several sophisticated computational packages such as PHCpack \cite{Ver:99}, PHoM \cite{GKKTFM:04}, Bertini \cite{SW:95} and HOM4PS2 \cite{L:03} which can be used to solve systems of multivariate polynomial equations. They are all available as freewares from the respective research groups. In the Appendix, we compare the Groebner basis technique and the NPHC method with technical remarks.

\subsection{Results}
We present the results by classifying the obtained solutions in terms of the number of positive and negative eigenvalues of the corresponding Hessian matrix, or the Faddeev-Popov operator, because then we can use the Neuberger zero as a necessary condition for having all the solutions.

\subsubsection{Anti-periodic Boundary Conditions}
We now explore the simplest non-trivial case in higher dimensional lattices which is the standard LLG functional on a $3\times 3$ lattice with the trivial orbit and anti-periodic boundary conditions
(i.e., the classical XY model) and also a random orbit case (i.e., randomly chosen $\phi_{i,\mu}\in (-\pi, \pi]$).
%(Special Random Orbit 3 and Random Orbit 3, in laptop)
For both cases, the CBB was $262144$ and the SMV was $148480$. For the trivial orbit case, the total number of real and complex solutions was $10738$ out of which there were $2968$ real solutions, i.e., Gribov copies. Similarly, for the random orbit case, the total number of real and complex solutions was $20558$ out of which there were $2480$ real solutions\footnote{Note that we do not intend to compare the efficiency of the available packages. However, to give an idea how computationally intensive such a calculation is, we note that the HOM4PS2 package took around $65$ and $120$ minutes, respectively, to run these systems on a Linux single-processor desktop machine.}.
It is important at this stage to mention a few specifics about these solutions. We give details for the trivial orbit case, which are similar to the other cases, below.
\begin{enumerate}
\addtolength{\itemsep}{-8pt}
 \item A solution means that a set of values of $s_{i}$'s and $c_{i}$'s satisfies each of the $18$ equations with tolerances $1\times 10^{-10}$. All the solutions come with real and imaginary parts. A solution is a real solution if the imaginary part of each of the $18$ variables (i.e., all $s_{i}$ and $c_{i}$) is less than or equal to the tolerance $1\times 10^{-6}$ (below which the number of real solutions does not change, i.e., it is robust for all the cases we consider in this discussion). The original trigonometric equations are satisfied with tolerance $1\times 10^{-10}$ after $s_{i}$ and $c_{i}$ are transformed back to $\theta_{i}$. All these solutions can be further refined with an \textit{arbitrary precision}. This is a remarkable success of the method because then these solutions are close to the \textit{exact solutions}.

\item The sum of the signs of the Faddeev-Popov determinants for all real solutions for the trivial and random orbit cases is $\sim 10^{-11}$, which is numerically zero, yielding the expected Neuberger zero as mentioned in the Introduction.

\item For the trivial orbit case, there are exactly $1152$ real solutions which have zero Faddeev-Popov determinant with tolerance $1\times 10^{-8}$. These solutions constitute the set of singular loci or the Gribov horizons. They can be further classified in terms of the  number of zero eigenvalues of the Faddeev-Popov operator at each of the solutions. This amounts to classifying the singular solutions of the polynomial system in terms of their multiplicities using the so-called deflation singularities technique\cite{OjWaMi83}. For both the random orbit cases, there is no Gribov horizon, i.e., all the solutions are non-singular.

 \item For the trivial orbit case, the remaining $1816$ real solutions
   have nonsingular Faddeev-Popov determinant, i.e., nonsingular
   solutions. All the nonsingular solutions of both the trivial and
   the random orbit cases can be classified by the number of negative
   eigenvalues (Table
   \ref{table:sllg_ap_bc_trivial_and_random_orbit_neg_e_values}),
   which shows the expected two-fold symmetry giving rise to the
   Neuberger zero.

%=================================================================

\begin{table}[t]
\begin{center}
\begin{tabular}{|r|r|r|r|r|r|}
   \hline
    boundary conditions & orbit & \multicolumn{3}{c}{real solutions} & sum of \\
%    boundary conditions & orbit & real solutions & sum of \\
            &           & total number & zero det. & non-zero det. & signs of det.\\
   \hline
   anti-periodic & trivial & 2968         & 1152      & 1816     &  0 \\
   \hline
            &      random & 2480      &   0      & 2480        &  0 \\
   \hline
    periodic & trivial & 270        & 52      &   218    & 0 \\
    \hline
            &      random & 224       &  0      & 224     & 0 \\
  \hline
  \end{tabular}
\parbox{11.4cm}{\caption {An overview on the number of real solutions
     for the different cases (orbits and boundary conditions). See Section $3$ for the detailed description.\label{table:summary_of_real_solns}}}
\end{center}
\end{table}

\begin{table}[t]
\begin{center}
\begin{tabular}{|r|r|r|r|r|r|r|r|r|r|r|r|}
\hline
Case & i & 0 & 1 & 2 & 3 & 4 & 5 & 6 & 7 & 8 & 9 \\
\hline
\mbox{antiperiodic b.c., trivial} & $K_{i}$ & 2 & 18 & 216 & 342 & 330 & 330 & 342 & 216 & 18 & 2 \\
\hline
\mbox{antiperiodic b.c., random} & $K_{i}$ & 2 & 58 & 202 & 402 & 576 & 576 & 402 & 202 & 58 & 2 \\
\hline
\mbox{periodic b.c., trivial} & $K_{i}$ & 8 & 28 & 38 & 42 & 56 & 38 & 7 & 1 & - & -\\
\hline
\mbox{periodic b.c., random} & $K_{i}$ & 2 & 12 & 30 & 61 & 72 & 38 & 8 & 1 & - & - \\
\hline
\end{tabular}
\parbox{11.4cm}{\caption{Summary of the number of real solutions $K_{i}$ with
    $i$ negative eigenvalues on the $3\times 3$ lattice, for the different cases (orbits and boundary conditions).\label{table:sllg_ap_bc_trivial_and_random_orbit_neg_e_values}}}

\end{center}
\end{table}

\end{enumerate}
See Table \ref{table:summary_of_real_solns} for the summary.

\subsubsection{Periodic Boundary Conditions}
After eliminating the constant zero mode for the $3\times 3$ lattice
case with periodic boundary conditions (by taking, say, two of the
$\theta_{i}$'s at the corners of the lattice to be $0$, without losing
generality. See Ref.~\cite{Mehta:2009} for more details), we have $14$
equations and $14$ variables. The number of solutions is $270$ out of
which $218$ are non-singular and $52$ are singular. The Faddeev-Popov operator is now a $7\times 7$
dimensional matrix. The sum of signs
of the Faddeev-Popov determinants is zero.
% The results are summarized in Table \ref{table:sllg_p_bc_trivial_orbit_pos_neg_e_values}.
%\begin{table}[h]
%\begin{tabular}{|r|r|r|r|r|r|r|r|r|r|r|r|r|}
%\hline
%$i$ & 0 & 1 & 2 & 3 & 4 & 5 & 6 & 7 \\
%\hline
%$P_{i}$ & 1 & 7 & 38 & 56 & 42 & 38 & 28 & 8 \\
%\hline
%$K_{i}$ & 8 & 28 & 38 & 42 & 56 & 38 & 7 & 1 \\
%\hline
%\end{tabular}\caption{Summary of the number of stationary points $P_{i}$ with $i$ positive eigenvalues, and the number of stationary points $K_{i}$ with $i$ negative eigenvalues for the $3\times 3$ lattice, trivial orbit, periodic boundary conditions.}\label{table:sllg_p_bc_trivial_orbit_pos_neg_e_values}
%\end{table}
Similarly, for a random orbit
% output_2dim_3x3_standard_periodic_theta13_theta33_being_zero_random_orbit.txt.changed
the number of real solutions is $224$ and again the sum of signs of
the Faddeev-Popov determinants is $0$.
 The results are summarized in Table
 \ref{table:sllg_ap_bc_trivial_and_random_orbit_neg_e_values}.

%\begin{table}
%\begin{tabular}{|r|r|r|r|r|r|r|r|r|r|r|r|r|}
%\hline
%$i$ & 0 & 1 & 2 & 3 & 4 & 5 & 6 & 7 \\
%\hline
%$P_{i}$ & 1 & 8 & 38 & 72 & 61 & 30 & 12 & 2 \\
%\hline
%$K_{i}$ & 2 & 12 & 30 & 61 & 72 & 38 & 8 & 1 \\
%\hline
%\end{tabular}\caption{Summary of the number of stationary points $P_{i}$ with $i$ positive eigenvalues and the number of stationary points $K_{i}$ with $i$ negative eigenvalues, for the $3\times 3$ lattice, random orbit, periodic boundary conditions.}\label{table:sllg_p_bc_random_orbit_pos_neg_e_values}
%\end{table}

\section{Summary}
By taking the standard lattice Landau gauge-fixing (LLG) functional (the classical Hamiltonian of the random phase XY model (RPXYM)) as an example, we showed that the non-linearity of the corresponding gauge-fixing equations (extremizing equations) is \textit{polynomial-like}. With additional constraint equations, the combined system of equations can be treated as a system of polynomial equations with all variables defined over $\mathbb{C}$. Though we are interested in only real solutions of these polynomial systems, we used Complex Algebraic Geometry concepts over Real Algebraic Geometry due to the stronger results available in the former.
To solve the corresponding polynomial systems exactly, one can use an elegant algorithm, called the Buchberger algorithm (see Appendix A). Though the algorithm works well for the one-dimensional lattice, for the $3\times 3$ lattice, it is beyond the capabilities of a standard desktop machine. However, the method can be useful in finding a parametrization of the SU($N$) matrices~\cite{car:09} and hence can then be used, for example, to extend the method of restricting a path integral of the SU($N$) Georgi-Glashow model to the 't~Hooft-Polyakov sectors by using the twisted C-periodic boundary conditions~\cite{Edwards:2009bw} for odd $N$ and to precisely define the modified lattice Landau gauge for the SU($3$) case.

For the two-dimensional case, we used the Numerical Polynomial Homotopy Continuation (NPHC) method, which gives \textit{all} solutions of a polynomial system numerically. This method does not suffer from the technical difficulties of the Buchberger algorithm or its variants, and in principle this method can find the solutions for those systems which may be intractable for the former method. By computing the eigenvalues of the Hessian matrix at these solutions, one can obtain the global minimum, up to machine precision, of a multivariate function whose extremizing equations can be translated to polynomial equations.

The $3\times 3$ anti-periodic boundary conditions system for the trivial and a random orbit can be solved using the NPHC method and that there is an exact cancelation of the signs of the Faddeev-Popov determinants in this case, i.e., the Neuberger zero. We classified all the solutions in terms of the number of negative and positive eigenvalues of the corresponding Faddeev-Popov operator. Apparently, the number of Gribov copies is orbit dependent, though the Neuberger zero is orbit independent. It would be important to compute the number of solutions for different random orbits for the $3\times 3$ lattice with the modified LLG using the NPHC method. Efforts to solving the corresponding gauge-fixing equations for the modified LLG for compact U($1$) and also the linear covariant gauge-fixing equations for compact U($1$), in the spirit of~\cite{Cucchieri:2009kk, Cucchieri:2008zx}, are in progress. It would also be very interesting to solve the Langevin dynamics equations for the XY model and other models ~\cite{Aarts:2008wh} using algebraic geometry methods.

\section{Acknowledgement}
DM would like to thank the developers of the PHCpack, HOM4PS2, Bertini, PHoM and Singular, and to Hirokazu Anai, Guillaume Moroz, Antonio Montes, Akira Suzuki, Vladimir Gerdt, Daniel Robertz, Anton Ilderton, Yang-Hui He, JM Kosterlitz, Jon-Ivar Skullerud, Teresa Mendes, Attilio Cucchieri and Paul Watts for their help and critical remarks on this work. DM was supported by the Science Foundation of Ireland and by the British Council Researcher Exchange Programme to visit the Imperial College London. Support by the Australian Research Council is also acknowledged.

\appendix

\section{Computational Algebraic Geometry}
Since many important and useful results in algebraic geometry are mainly developed for the complex field $\mathbb{C}$, we treat our polynomial equations as having complex variables, and in the end we use only real solutions and throw the remaining solutions away (see, e.g., ~\cite{Gray:2008zs}). One can then use an important result from computational algebraic geometry: after specifying an ordering of the monomials of the system (denoted as $x\succ y \succ z$ for the ordering in which $x$ is placed prior to $y$ and $y$ is placed prior to $z$), one can transform a given system of multivariate polynomial equations, referred to as an \textit{ideal}, to another one which has the same solutions but is easier to solve. This new system is called a Groebner basis for the given monomial ordering.

There is a well defined procedure to find a Groebner Basis for any given ideal and monomial order, called the Buchberger algorithm. It should be noted that the Buchberger algorithm reduces to Gaussian elimination in the case of linear equations, i.e., it is a generalization of the latter. Similarly it is a generalization of the Euclidean algorithm for the computation of the Greatest Common Divisors of a univariate polynomial. Recently, more efficient variants of the Buchberger algorithm have been developed to obtain a Groebner basis, e.g., F4, F5 and Involution Algorithms~\cite{gerdt-98}. Symbolic computation packages such as Mathematica, Maple, Reduce, etc., have built-in commands to calculate a Groebner basis for a given monomial. Singular, COCOA and McCauley2 are specialized packages for Groebner basis and Computational Algebraic Geometry, available as freeware. MAGMA is also such a specialized package available as a non-free package. Rather than going into the details of the specifics and technicalities of this algorithm, we dive into the practical applications of the Groebner basis technique relevant to our problem and refer the reader to the above mentioned references for further details.

\subsection{The Groebner Basis Technique At Work}
Here, we provide a practical example of how the Groebner basis technique can be used for our systems. Firstly, for the polynomials for the one-dimensional lattice with three lattice sites, anti-periodic boundary conditions and the trivial orbit case, the corresponding ideal (denoted as the polynomials put between '$<$' and '$>$') is
\begin{eqnarray}
I_{3} &=& <-c_{2} s_{1} - c_{3} s_{1} + c_{1} s_{2} - c_{1} s_{3}, c_{2} s_{1} - c_{1} s_{2} - c_{3} s_{2} + c_{2} s_{3}, -c_{3} s_{1} + c_{3} s_{2} - c_{1} s_{3} - c_{2} s_{3},\nonumber \\
& & c_{1}^{2} + s_{1}^{2} - 1, c_{2}^{2} + s_{2}^{2} - 1, c_{3}^{2} + s_{3}^{2} - 1 >. \label{eq:ideal_antip_standard_trivial_one_dim}
\end{eqnarray}
For the ideal in Eq.~(\ref{eq:ideal_antip_standard_trivial_one_dim}), a Groebner basis for the so-called lexicographic ordering $c_{1}\succ
 s_{1}\succ c_{2}\succ s_{2}\succ c_{3}\succ s_{3}$ is
\begin{eqnarray}
G_{3} &=& < -s_{3} + s_{3}^3, c_{3} s_{3},-1 + c_{3}^{2} + s_{3}^{2}, -s_{2} + s_{2} s_{3}^{2}, c_{3} s_{2}, s_{2}^{2} - s_{3}^{2}, c_{2} s_{3}, c_{2} s_{2},\nonumber \\
& &  -1 + c_{2}^{2} + s_{3}^{2}, -s_{1} + s_{1} s_{3}^{2}, c_{3} s_{1}, c_{2} s_{1}, s_{1}^{2} - s_{3}^{2}, c_{1} s_{3} , c_{1}s_{2}, c_{1} s_{1}, -1 + c_{1}^{2} + s_{3}^{2} > .
\end{eqnarray}
As noted earlier, the solutions of this system are the same as the
 original system. Here, the first equation in $G_{3}$ is a univariate polynomial
 in variable $s_{3}$ and solving it is simple because it
 can be factorized as $s_{3}(s_{3}^{2} - 1) = 0$ giving $s_{3} = 0$ and $s_{3} = \pm 1$. Using this and \textit{back-substitution}, all the solutions can be obtained and are as follows:
\begin{equation}
(c_{1},c_{2},c_{3},s_{1},s_{2},s_{3}) = \lbrace (0,0,0,\pm 1,\pm 1,\pm1), (\pm 1,\pm 1,\pm 1,0,0,0)\rbrace .\label{eq:sol_c_s_one_dim_3_antip_standard_trivial}
\end{equation}
Thus, the solution space of the system $G_{3}$, called the \textit{variety} of $G_{3}$ and denoted as $V(G_{3})$, is the above mentioned
 set of $16$ isolated points in a $6$-dimensional affine space.
This is the known result from Refs.~\cite{vonSmekal:2007ns,vonSmekal:2008es,Mehta:2009} in terms of $\theta_{i}$'s, i.e.,
\begin{equation}
(\theta_{1},\theta_{2},\theta_{3}) = \lbrace\mbox{all $2^{3}$ permutations of $0/\pi$}\rbrace \cup\mbox{ } \lbrace\mbox{all $2^{3}$ permutations of $-\frac{\pi}{2}/\frac{\pi}{2}$}\rbrace .
\end{equation}
All of these solutions are correctly reproduced in Eq.~(\ref{eq:sol_c_s_one_dim_3_antip_standard_trivial}). Similar computation can be performed for the periodic boundary conditions case~\cite{Mehta:2009}.

For an ideal which is known to have only isolated solutions (called a zero-dimensional ideal) with a lexicographic ordering one can always find a Groebner basis in an \textit{upper diagonal} form, analogous to the Gaussian Elimination method, such that at least one polynomial is univariate with the others having an increasing number of variables.

Interpreting the gauge fixing equations in terms of Algebraic Geometry allowed us to deal with the \textit{actual non-linearity} of the equations, rather than treating them as linear equations as done in Refs.~\cite{Mehta:2009,vonSmekal:2007ns,vonSmekal:2008es}. Specifically, in this interpretation the corresponding method does not make any distinction between the equations arising from a one-dimensional lattice or those arising from a higher dimensional lattice. In theory, as long as one can obtain a Groebner basis, the equations can be exactly solved. For the one-dimensional case, it was easy to go up to a fairly big number of lattice sites on a single machine. However, in general, obtaining a Groebner basis is very difficult due to an algorithmic complexity, known as Exponential Space complexity, which roughly means that the RAM required by the computation blows up exponentially. In particular, on a regular single desktop machine with $2$ GB RAM, we could not obtain a Groebner basis for the trivial orbit case (i.e., the classical XY model) and with anti-periodic boundary conditions for a $3\times 3$ lattice, using Singular3.2. The corresponding system is made of $18$ equations, each of degree $2$, in $18$ variables. However, a more powerful machine should be able to obtain a Groebner basis for these systems.

Recently, V.~Gerdt and Daniel Robertz were able to compute a Groebner Basis for this system over $F_{2}=\{0,1\}$ using a Linux machine with AMD Opteron (TM) processor $285$ $2600$ MHz, $4$ cores, $16$ GB memory and with Magma (V2.14.14) in less than $10$ hours with degree reverse lexicographic ordering~\cite{gerdt_robetz_pvt_communication}. This is a very important step towards solving the system. However, since the computation is over $F_{2}$ not all the solutions over $\mathbb{C}^{18}$ can be obtained.

\subsection{Comparison between the Groebner basis technique and the NPHC Method}
The NPHC method is strikingly different from the Groebner basis technique in that the algorithm for the former suffers from no known major complexities. Moreover, the path tracking is \textit{embarrassingly  parallelizable}, because all the start solutions can be tracked completely independently of each other. This feature along with the rapid progress towards the improvements of the algorithms makes the NPHC well suited for many physical problems arising in condensed matter theory, lattice QCD, etc.

For most systems of polynomial equations in practice, we do not know the actual number of solutions from the beginning. So even after obtaining all solutions through the homotopy continuation method, we would still prefer to have some kind of verification of the solutions. In the standard LLG fixing case, we used the Neuberger zero as a necessary condition. However, this is certainly not a sufficient condition and further checks may be required for bigger systems. This is called \textit{certifying} the solutions, which has been recently developed using the so-called Tropical Algebraic Geometry~\cite{adrovic-2008}. We anticipate that our results mentioned above will serve as ideal test systems for all such new ideas.

To solve the corresponding equations for the periodic boundary conditions case, another method is to leave the constant zero modes in the system and use the Numerical Algebraic Geometry (NAG)~\cite{SVW:96}, which is the generalization of the NPHC method to positive-dimensional ideals.

Recently, a computational Algebraic Geometry approach to find vacuum configurations in string phenomenology (where mathematically the problem reduces to minimization of the so-called superpotentials which exhibit polynomial-like non-linearity) is proposed in Refs.~\cite{Gray:2008zs, Gray:2007yq}. A very efficient Mathematica package, called STRINGVACUA, has been developed with an interface with the computer algebra system Singular. This package relies on the Groebner basis technique in solving the extremizing polynomial equations. It is anticipated that incorporating the NPHC method can extend the applicability of the package and the approach beyond its present status.

It should be noted that there are several promising methods such as the Discriminant Variety~\cite{Lazard:04} and finding real solutions out of complex curves~\cite{Lu06findingall} from computational real algebraic geometry which are emerging as alternatives to the above mentioned methods in some of the important problems mentioned in the Introduction.

\end{document}